\documentclass[twocolumn,preprintnumbers,prd,superscriptaddress,nofootinbib,floatfix,showpacs,showkeys]{revtex4-1}
\usepackage{amsmath}
\usepackage{amsfonts}
\usepackage{amssymb}
\usepackage{graphicx}
\usepackage{color}
\usepackage{hyperref}
\usepackage{booktabs}
\usepackage{hyperref}
\usepackage{cleveref}
\usepackage{braket}
\usepackage{mathtools}
\usepackage[rightcaption]{sidecap}
\usepackage{subfigure}
\usepackage{soul}
\usepackage{enumitem}
\usepackage{comment}
\usepackage{multirow}

\definecolor{vividviolet}{rgb}{0.62, 0.0, 1.0}
\definecolor{amaranth}{rgb}{0.9, 0.17, 0.31}
\definecolor{palatinateblue}{rgb}{0.15, 0.23, 0.89}
\definecolor{brightpink}{rgb}{1.0, 0.0, 0.5}
\definecolor{cornflowerblue}{rgb}{0.39, 0.58, 0.93}
\definecolor{deepcarminepink}{rgb}{0.94, 0.19, 0.22}
\definecolor{radicalred}{rgb}{1.0, 0.21, 0.37}
\hypersetup{ linktoc=all,
    colorlinks, linkcolor={palatinateblue},
    citecolor={brightpink}, urlcolor={amaranth}
}

\newcommand{\be}{\begin{equation}}
\newcommand{\ee}{\end{equation}}
\newcommand{\bs}{\begin{split}}
\newcommand{\bea}{\begin{eqnarray}}
\newcommand{\eea}{\end{eqnarray}}

\newcommand{\bes}{\begin{subequations}}
\newcommand{\ees}{\end{subequations}}

\begin{document}

\title{Temperature definitions and phase transitions within non-minimal large and small inflationary potentials}

\author{Jesus Anaya-Galeana}
\email{jesus.anaya@correo.nucleares.unam.mx}
\affiliation{Instituto de Ciencias Nucleares, Universidad Nacional Aut\'onoma de M\'exico, AP 70543, M\'exico, DF 04510, Mexico.}

\author{Orlando Luongo}
\email{orlando.luongo@unicam.it}
\affiliation{School of Science and Technology, University of Camerino, Via Madonna delle Carceri, Camerino, 62032, Italy.}
\affiliation{Istituto Nazionale di Fisica Nucleare (INFN), Sezione di Perugia, Perugia, 06123, Italy.}
\affiliation{SUNY Polytechnic Institute, 13502 Utica, New York, USA.}
\affiliation{INAF - Osservatorio Astronomico di Brera, Milano, Italy.}
\affiliation{Al-Farabi Kazakh National University, Al-Farabi av. 71, 050040 Almaty, Kazakhstan.}

\author{Hernando Quevedo}
\email{quevedo@nucleares.unam.mx}
\affiliation{Instituto de Ciencias Nucleares, Universidad Nacional Aut\'onoma de M\'exico, AP 70543, M\'exico, DF 04510, Mexico.}
\affiliation{Dipartimento di Fisica and Icra, Universit\`a di Roma “La Sapienza”,
Piazzale Aldo Moro 5, I-00185 Roma, Italy.}

\begin{abstract}
We explore and compare two distinct temperature definitions for scalar field inflation in the context of small- and large-field potentials. The first is based on a real gas, fluid-like temperature, $T_{RG}$, while the second corresponds to a relativistic species-like temperature, $T_{RS}$. We derive the fundamental thermodynamic relations for both and analyze their implications for the most viable inflationary potentials, consistent with Planck constraints. We also investigate non-minimally coupled scenarios, finding that $T_{RS}$ is the most self-consistent choice, as it decreases during inflation, satisfies standard thermodynamic laws, and exhibits frame-independent behavior in both the Jordan and Einstein frames. Remarkably, the $T_{RS}$ approach shows that the inflaton’s dynamics is well-described by Van der Waals-like isotherms, linking inflationary evolution to thermodynamic phase transitions. We find that the onset of inflation is associated with a phase transition acting as the ``trigger'' of the inflationary epoch. Our analysis highlights inconsistencies in the hilltop potential and, more generally, in small-field potentials unless a non-minimal coupling is introduced. Conversely, the Starobinsky and $\alpha$-attractor models emerge as the most suitable paradigms. We further show that \emph{frame independence} is achieved only for coupling values $\zeta \leq 1/6$, supporting very small values. Finally, our study of natural inflation with non-minimal coupling reveals a strong dependence on the coupling parameter, where bounds associated with thermodynamic phase transitions coincide with observationally viable ranges, suggesting that thermodynamic considerations may provide an additional criterion to discriminate among inflationary scenarios.
\end{abstract}



\maketitle
\tableofcontents

\section{Introduction}

Thermodynamics plays a crucial role in cosmology, the study of the origin, evolution, and ultimate fate of the universe. The laws of thermodynamics provide a framework for understanding the behavior of energy, matter, and entropy on cosmic scales. For example, during Big Bang nucleosynthesis, the formation of light elements—such as hydrogen, helium, and traces of lithium—in the first few minutes of the universe was governed by thermodynamic equilibrium \cite{grohs2023bigbangnucleosynthesis,RevModPhys.88.015004,IOCCO20091}. Similarly, during the recombination epoch, the universe cooled sufficiently for electrons to combine with protons, forming neutral hydrogen and allowing photons to travel freely. This process led to the creation of the cosmic microwave background (CMB) \cite{wong2008cosmologicalrecombination}.

However, for thermodynamics to be applicable, a well-defined temperature is required. Throughout cosmic history, it has been assumed that the universe's content follows a statistical distribution, enabling the definition of temperature.
In the context of scalar field inflation, temperature is often introduced as a parameter through the concept of a thermal state or thermal ensemble, implying (at least local) thermal equilibrium. Cosmologically, temperature is often incorporated into an effective potential of the form $ V_{eff}(\phi,T) = V_0(\phi) + V_T(\phi,T) $, where $V(\phi)$ is the zero temperature potential and $V_T(\phi, T)$ is the thermal correction to the potential. $T$ here is the temperature of the thermal bath surrounding the scalar field.
The thermal corrections arise from the contributions of thermal fluctuations to the effective potential. For a scalar field, the thermal correction depends on the mass of the scalar field and its quantum nature (i.e. bosonic or fermionic). A rigorous definition of temperature for a scalar field in cosmology remains an open question.
During inflation, the universe rapidly departs from thermal equilibrium due to the exponential dilution of particle number densities\footnote{Interactions become too infrequent to maintain equilibrium.}. If we assume thermal equilibrium at onset of inflation, post-inflation particles may retain a thermal distribution, allowing an effective temperature definition analogous to the CMB temperature even without equilibrium.

A key challenge is defining a temperature function that can be used to characterize phase transitions associated with scalar fields. Such a definition would enable for a thermodynamic description of the scalar field itself, particularly during reheating. This prerogative would yield the following points.
\begin{itemize}
    \item[-] Determining the thermodynamic entropy evolution through reheating and the radiation-dominated epoch.
    \item[-] Defining an effective temperature and pressure equation of state (EoS), extending the simplistic assumption of having radiation and/or pressureless fluids \cite{PhysRevLett.119.061301,PhysRevD.73.023501,Mu_oz_2015,Saha_2020}.
\end{itemize}
Previous approaches, e.g. Refs.   \cite{PhysRevD.45.3429,PhysRevD.47.2302}, proposed a temperature definition that matches the behavior of the inflationary and radiation eras and can be expressed in terms of the scale factor. Notably, this definition is independent of the inflation mechanism. However, alternative views exist, i.e., treating the temperature EoS as a real fluid and, moreover, mimicking an ideal gas in the radiation-dominated limit.

Motivated by the above considerations, we therefore explore two possible thermodynamic frameworks for the inflaton field as reported below.
\begin{enumerate}
    \item A real gas fluid-like temperature: Models the inflaton as a real perfect fluid, yielding distinct dynamics.
    \item A relativistic species-like temperature: A generalization of the temperature defined for relativistic particles through the calculus of energy density for relativistic species via the distribution function, compatible with scalar-field phase transitions.
\end{enumerate}

In this respect, using the slow-roll approximation, we derive an EoS for pressure in terms of energy density and a thermodynamic volume function\footnote{Ensuring that the cosmic volume scales as $\propto a^3$ enables to rephrase our treatment in terms of the scale factor. More complicated scalings, such as in terms of the apparent horizons, see e.g. Ref. \cite{Cai:2006rs}, go beyond the purposes of this paper.}. From this, we obtain a fundamental equation for entropy, expressible via energy, volume, scale factor, or the inflaton field. We then analyze the following main steps.
\begin{itemize}
    \item[-] Isotherms and stability conditions.
    \item[-] Criteria for thermodynamic phase transitions and critical points.
\end{itemize}

In this work, we test six minimally coupled potentials and one non-minimally coupled case. Our results indicate that the relativistic species-like temperature decreases during inflation and exhibits phase transitions and critical points consistent with scalar-field theory. In contrast, the real gas temperature increases during inflation, which contradicts thermodynamic expectations. The thermodynamic approach enables a systematic classification of models based on properties such as phase transitions and critical points. We find that phase transitions are a key characteristic of observationally favored models like Starobinsky and $\alpha$-attractors, while critical points are associated with more constrained models such as hilltop inflation. This thermodynamic behavior provides an additional criterion for model classification. For the non-minimally coupled scenario, we determine that the coupling parameter $\zeta$ has a maximum value of $1/6$ to maintain \emph{frame-invariant} thermodynamics.

The paper is structured as follows. In Sect. \ref{section 2}, we discuss the dynamics of single-field inflation within the slow-roll approximation for both minimally and non-minimally coupled fields. Additionally, we derive the EoS for the pressure in each scenario. In Sect. \ref{section3}, we introduce two distinct notions of temperature for inflationary fields and develop the thermodynamics associated with each. Sect. \ref{section4} analyzes six different potentials for single scalar field inflation, focusing on their thermodynamics, particularly critical points and phase transitions. We also compare these thermodynamic results with the conventional phase transitions observed in scalar field theory. Finally, Sect. \ref{section5} presents the conclusions and perspectives of our work, highlighting that one of the studied temperature definitions aligns with the expected behavior of scalar field potentials.

\section{Inflationary set up} \label{section 2}

The inflationary set up was usually first described by \emph{old inflation}, that consists in the overall treatment to accelerate the very early universe by virtue of a further scalar field, dubbed \emph{inflaton}, bypassing a symmetry breaking mechanism that induces the inflationary stage. In the \emph{new inflation} picture, in contrast, the mechanism of acceleration is recovered but does not undergo a direct phase transition, preferring a chaotic inflationary potential, guaranteeing a strong speed up without formation of unexpected bubbles, \emph{de facto}, the most serious problem associated with  old inflation \cite{PhysRevD.23.347,GUTH1983321,1983veu..conf..251S}.

In both scenarios, inflation can be minimally or non-minimally coupled with external fields \cite{1981ZhPmR..33..549M,bezrukov2008nonminimalcouplinginflationinflating,Hertzberg_2010,PhysRevD.41.1783,Kodama_2022,Gao:2025onc}. Examples are couplings with spectators \cite{2013JCAP...10..042K,Dimopoulos_2018,Chen:2009zp}, environment or thermal fields \cite{Berera_1995,Moss_2008,Samart_2022,Anber_2010,KIEFER_1998}, curvature scalar \cite{Bauer_2011,J_rv_2017}, and so on. Quite interestingly, the Jordan and Einstein frames problem is  particularly conceivable here \cite{PhysRevD.75.023501,sym12010147,PhysRevD.57.811,PhysRevD.62.044042,PhysRev.125.2163,PhysRevD.50.5039,faraoni1998conformaltransformationsclassicalgravitational,Flanagan:2004bz,BERTOLAMI1987161}. For example, postulating a minimally coupled Lagrangian in the Jordan frame is equivalent to handling a non minimally coupled Lagrangian in the Einstein frame, starting from its non-minimal version written in the Jordan frame; see e.g.  \cite{figueroa2024dynamicsnonminimallycoupledscalar,Bezrukov_2011,PhysRevD.76.084039,faraoni1998conformaltransformationsclassicalgravitational}.

The two possibilities are, however, not clearly matchable, and possibly the non equivalence of frames can persist even at the level of particle production, see e.g. \cite{belfiglio2024comparinggeometricgravitationalparticle,2023PhRvD.108h3535T,PhysRevD.61.103501,2023PDU....3901169Z,2022JCAP...02..029C,2024PDU....4401458B,2025arXiv250216262Z}.

Accordingly, an in-depth study of inflationary dynamics adopting a scalar field Lagrangian in one frame may be sufficient to describe the overall dynamics, recalling moreover that relativistic fields are the basic demands to guarantee inflation, since there is no need of handling quantum inflationary fields, as demonstrated in Refs. \cite{Luongo_2024,Hertzberg_2010,PhysRevD.90.103516, Capozziello:1996xg}.

In what follows, we analyze both minimally and non-minimally coupled frameworks, studying the inflationary thermodynamics in both the Jordan and Einstein frames.

\subsection{Minimally coupled scalar field}

Within the inflationary paradigm, the minimally-coupled scalar field  action reads \cite{LINDE1982389,LINDE1983177,Linde2007,PhysRevD.50.7222,martin2003inflationprecisioncosmology}
    \begin{equation}\label{eq: accion campo escalar}
    S = \int d^4 x \sqrt{-g} \left[R -\frac{1}{2}\partial_\mu \phi \partial^\mu \phi - V(\phi) \right] \, .
\end{equation}
where $V(\phi)$ is the inflationary potential.

Varying the action, one obtains the Klein-Gordon equation of motion for the inflaton, $\phi$, and the Einstein equations, $G_{\mu \nu} = R_{\mu \nu} - \frac{1}{2}g_{\mu \nu}R = 8\pi G_N T_{\mu \nu}$, fueled by the energy-momentum tensor $T_{\mu\nu}$. Thus, we write
\begin{subequations}
\begin{align}
    &\Ddot{\phi} + 3H \Dot{\phi} - \frac{\nabla^2 \phi}{a^2} + V'(\phi) = 0 ,\label{eq: movimiento inflaton}\\
    &T_{\mu \nu} = - \frac{2}{\sqrt{-g}} \frac{\delta S}{\delta g^{\mu \nu}} = \partial_{\mu} \phi \partial_{\nu} \phi - g_{\mu \nu}  \left[ \frac{1}{2}\partial_\mu \phi \partial^\mu \phi + V(\phi) \right] \, .\label{eq: tensor energia campo escalar}
\end{align}
\end{subequations}

Consequently, one immediately can notice that scalar fields and barotropic fluids are not the same constituents, exhibiting profound differences \cite{MSMadsen_1988,Unnikrishnan:2010ag,DIEZ_TEJEDOR_2005,PhysRevD.80.124040}.

However, limiting the kinetic sign, there are approaches that allow us to frame scalar fields into perfect fluids \cite{PhysRevD.85.024040,article,Arroja_2010}.
This interpretation allows one to define an energy density and pressure as
\begin{subequations}
\begin{align}
    &\rho_{\phi} = {T_{0}}^0 = \frac{\Dot{\phi}^2}{2} + V(\phi) + \frac{ {( \nabla \phi )}^2}{6a^2} , \\
    &P_{\phi} = \frac{ {T^i}_i}{3} = \frac{\Dot{\phi}^2}{2} - V(\phi) - \frac{ {( \nabla \phi )}^2}{6a^2}.
    \end{align}
\end{subequations}

Conveniently, in a homogeneous and isotropic background, the overall inflation description reduces to
\begin{subequations}
    \begin{align}
    &\rho_{\phi} = \frac{\Dot{\phi}^2}{2} + V(\phi),\label{eq: densidad inflacion campo}\\
    &P_{\phi} =  \frac{\Dot{\phi}^2}{2} - V(\phi),\label{eq: presion inflacion campo}\\
    &\Ddot{\phi} + 3H\Dot{\phi} + V'(\phi) = 0,\\
    &H^2 =  \frac{1}{3M_{pl}^2} \left[ V(\phi) + \frac{1}{2}\Dot{\phi}^2 \right] .    \label{eq: friedmann with field}
    \end{align}
\end{subequations}
where $M_{pl}$ is the Planck mass. Now, defining the number of e-foldings as $N \equiv \ln (a/a_i)$, we definitely obtain
\begin{subequations}
\begin{align}
    &\frac{1}{3-\epsilon_1}\frac{d \phi^2}{d N^2} + \frac{d \phi}{d N} = - M_{pl}^2 \frac{d \ln V}{d \phi},\label{eq: field evolution}\\
    &\epsilon_1 \equiv \frac{1}{16\pi G_N} {\left( \frac{\partial_{\phi}V}{V} \right) }^2 = \frac{M_{pl}^2}{2} {\left( \frac{\partial_{\phi}V}{V} \right) }^2.    \label{eq: epsilon_1}
\end{align}
\end{subequations}
Eq. (\ref{eq: field evolution}) can be further reduced in the slow-roll approximation, yielding
\begin{equation}
    \label{eq: field evolution slow-roll}
    \Delta N = N - N_{i} = - \frac{1}{M_{pl}^2} \int_{\phi_i}^{\phi}\frac{V(\chi)}{\partial_{\chi}V(\chi)} d\chi .
\end{equation}
Inflation in the slow-roll approximation is well-defined as long as $\epsilon_1 \ll 1$, when $\epsilon_1 = 1$, inflation ends, setting a value for the scalar field at the end of inflation.

As stated above, the perfect fluid approximation is only partially accounted for scalar fields. However, let us note that we can combine Eqs. (\ref{eq: presion inflacion campo}) and (\ref{eq: densidad inflacion campo}) furnishing the weak energy condition $\Dot{\phi}^2\equiv P+\rho$ \cite{wald2010general}.

In the slow roll domain, we roughly have
$H \approx \frac{1}{M_{pl}\sqrt{3}}\sqrt{V(\phi)}$, so the time derivative of Eq. (\ref{eq: field evolution slow-roll}) gives
\begin{equation}
    \label{eq: phi punto}
    \Dot{\phi} = -\sqrt{\frac{V(\phi)}{3}} \left[ M_{pl}\frac{d \left(N_f - N\right)}{d\phi}\right]^{-1} .
\end{equation}
We have explicitly used the number of e-foldings to fix the sign of $\Dot{\phi}$, or equivalently, in order to avoid degeneration in $\Dot{\phi}$ due to its sign, we used the number of e-foldings,
reformulating the corresponding EoS for the pressure as
\begin{equation}
    \label{eq: Pressure EoS}
    P = - \rho + \frac{M_{pl}^2}{3} \left[ \frac{d V(\phi)}{d\phi}\right]^{2}\frac{1}{V(\phi)} .
\end{equation}

The second term of Eq. (\ref{eq: Pressure EoS}) is a function of the inflaton field; however, since the thermodynamic volume scales as $\propto a^3$ and because the number of e-foldings, being a function of the inflaton, one can always change the variable so that the second term is a function of the volume.

\subsection{Non-minimally coupled scalar field}

We may now consider the following action
\begin{equation}
    \label{eq: accion campo escalar nonmin}
    \begin{split}
    S &= \int d^4 x \sqrt{-g} \left[\frac{M_{pl}^2}{2}\left(1 - \frac{\zeta\phi^2}{M_{pl}^2}\right)R -\frac{1}{2}\partial_\mu \phi \partial^\mu \phi - V(\phi) \right],
    \end{split}
\end{equation}
where $\zeta$ is the coupling constant, usually ensured small enough not to alter the gravitational constant measurements \cite{PhysRevD.39.399,PhysRevD.61.103501}.

Here, we obtain the field equation for the inflaton $\psi$ and the energy-momentum tensor, respectively
\begin{subequations}
\begin{align}
    &\Ddot{\phi} + 3H \Dot{\phi} - \frac{\nabla^2 \phi}{a^2} + \frac{d V(\phi)}{d \phi} = -\zeta R\phi, \label{eq: movimiento inflaton nonmin}
    \\
    T_{\mu \nu} &=\partial_{\mu} \phi \partial_{\nu} \phi - g_{\mu \nu}  \left[\frac{1}{2}\partial_\mu \phi \partial^\mu \phi + V(\phi) \right]    \notag\\
    & +  \zeta \left[ G_{\mu\nu} + g_{\mu\nu}g^{\alpha\beta}\nabla_{\alpha}\nabla_{\beta} - \nabla_{\mu}\nabla_{\nu} \right] \phi^2.\label{eq: tensor energia campo escalar nonmin}
\end{align}
\end{subequations}

Within the homogeneity and isotropy context, we can compare the energy-momentum tensor of this theory with that of a perfect fluid. Doing so allows us to define the energy density and pressure as follows
\begin{equation}
    \label{eq: densidad inflacion campo nonmin}
    \rho_{\phi} = \frac{\Dot{\phi}^2}{2} + V(\phi) + 3\zeta H^2\phi^2 + 6\zeta H \phi \Dot{\phi} .
\end{equation}
\begin{equation}
    \label{eq: presion inflacion campo nonmin}
    \begin{split}
    P_{\phi} =& \, \,  \frac{\Dot{\phi}^2}{2} - V(\phi) + \zeta H^2\phi^2 + 8\zeta H \phi \Dot{\phi} \\
    & + 2\zeta\phi\Ddot{\phi} + 2\zeta\Dot{\phi}^2 - \frac{\zeta}{3}R \Dot{\phi}^2 .
    \end{split}
\end{equation}

It is important to note that neither the energy density nor the pressure are strictly positive (or negative) functions. This is because curvature explicitly influences their definitions through the Hubble function $H$ or directly.

From the second Friedmann equation, one can express the scalar curvature in terms of the field and its derivatives, having
\begin{subequations}
    \begin{align}
    R &= \frac{l(\phi)}{M_{pl}^2}\left[ (1-6\zeta)\left(\partial^{\mu}\phi\partial_{\mu}\phi\right) + 4V(\phi) - 6\zeta\phi\frac{\partial V}{\partial\phi} - T_m \right],      \label{eq: Ricci scalar nonmin}   \\
    l(\phi) &\equiv \left[1+1(6\zeta-1)\frac{\zeta \phi^2}{M_{pl}^2}\right]^{-1}, \label{eq: l nonmin}
    \end{align}
\end{subequations}
where $T_m$ is the trace of the matter energy-momentum tensor, that may be different or equal to zero.

In the Jordan frame, there also exist slow-roll conditions \cite{Kar_iauskas_2022, AKIN2020100691, Flanagan:2004bz}, which allow us to write
\begin{subequations}
    \begin{align}
    H^2 &\cong \frac{V(\phi)}{3M_{pl}^2f(\phi)}, \label{eq: aprox H nonmin}\\
    3H\Dot{\phi} &\cong \frac{2V(\phi)\frac{\partial f}{\partial\phi} - f(\phi) \frac{\partial V}{\partial \phi}}{f^2(\phi)},   \label{eq: aprox dotpsi nonmin}   \\
    \Dot{\phi}^2 &\ll 2H^2 M_{pl}^2 f(\phi), \\
    f(\phi) &= 1 - \frac{\zeta\phi^2}{M_{pl}^2}. \label{eq: f accion nonmin}
    \end{align}
\end{subequations}
The number of e-foldings for this frame is still defined as $\Delta N = \int Hdt$, however, it can be set to be a function of the field, by the use of the slow-roll conditions:
\begin{equation}
    \label{eq: field evolution slow-roll nonmin}
    \begin{split}
    \Delta N = N - N_{i} &= \int_{\phi_i}^{\phi} H(\chi) (\Dot{\chi})^{-1} d\chi, \\
    &= \frac{1}{M_{pl}^2} \int_{\phi_i}^{\phi}\frac{f(\chi)+6\zeta^2\frac{\chi^2}{M_{pl}^2}}{2f(\chi)\frac{\partial f}{\partial\chi} - f^2(\chi)\frac{1}{V(\chi)} \frac{\partial V}{\partial \chi}} d\chi .
    \end{split}
\end{equation}

Here, a slow-roll parameter also sets the ends of inflation as in the minimally coupled case. i.e., inflation ends at some value of the field that makes $\epsilon_{1,Jordan} = 1$, where
\begin{equation}
    \label{eq: slow-roll parameter jordan frame}
    \epsilon_{1,Jordan} = \frac{1}{2}\left[\frac{f^2(\phi)M_{pl}^2}{f(\phi)+6\zeta^2\frac{\phi^2}{M_{pl}^2}}\right]\left[\frac{1}{V(\phi)}\frac{\partial V}{\partial \phi} + \frac{4\zeta}{f(\phi)}\frac{\phi}{M_{pl}^2}\right]^2 .
\end{equation}

Again, the weak energy condition yields
\begin{equation}
    \begin{split}
    P_{\phi}+\rho_\phi =&\Dot{\phi}^2\left[1-\zeta\left(\frac{R(\phi)}{3}+2\right)\right] \\
    &+ \zeta\phi\left[4H^2\phi + 14 H \Dot{\phi} + 2\Ddot{\phi} + 2\zeta\Dot{\phi}^2 \right].
    \end{split}
\end{equation}
Using the generalized slow-roll conditions, we can neglect the acceleration terms for the field. From Eqs. (\ref{eq: aprox H nonmin}, \ref{eq: aprox dotpsi nonmin}), we can express $H=H(\phi)$ and $\Dot{\phi} = \Dot{\phi}(\phi)$. In so doing, assuming no matter content, $T_m = 0$, we obtain the pressure EoS in terms of the energy density and a function of $\phi$. A quick manipulation yields\footnote{The goal is to express the function $F$ in terms of the thermodynamic volume $V_{th}$. We achieve this by relating $V_{th}$ to the number of inflationary e-foldings $\Delta N$ via $V_{th} = V_f \exp(-3\Delta N)$, where $V_f$ is the volume at the end of inflation. Since the field $\phi$ also evolves with $\Delta N$, we can combine these relationships to write $F$ as a function of $\phi(V_{th})$.}.
\begin{equation}
    \label{eq: Pressure EoS nonmin}
    P = - \rho + F(\phi),
\end{equation}
with
\begin{equation}
    \label{eq: F(phi) nonmin}
    \begin{split}
    F(\phi) &= (1+2\zeta)\frac{\left(2V(\phi)\frac{\partial f}{\partial\phi} - f(\phi) \frac{\partial V}{\partial \phi}\right)^2}{3V(\phi)f^3(\phi)} M_{pl}^2 \\
    & + \frac{14}{3}\zeta\phi \frac{2V(\phi)\frac{\partial f}{\partial\phi} - f(\phi) \frac{\partial V}{\partial \phi}}{f^2(\phi)} \\
    & + \frac{4}{3}\frac{V(\phi)}{f(\phi)} \frac{\zeta \phi^2}{M_{pl}^2} - \frac{\zeta}{3}\phi^2 R(\phi) .
    \end{split}
\end{equation}

As it will be clear later, the definitions of temperature, entropy, pressure, and other thermodynamic quantities depend on the function $F(V_{th})$, determined by comparing the energy-momentum tensor of the scalar field with that of a perfect fluid. Consequently, the thermodynamics of the system is based on the form of $F(V_{th})$. If $F(V_{th})$ is altered, the system's behavior will change accordingly.

Thus, the thermodynamics of a minimally and non-minimally coupled scalar field will be different. Summarizing, the pressure EoS for the inflaton is
\begin{equation}
    P = - \rho + F(V_{th})
\end{equation}
where $V_{th}\propto a^3$ is related to the inflaton field via the number of e-foldings (\ref{eq: relationship dN V})
\begin{equation}
    \label{eq: relationship dN V}
    \Delta N = \frac{1}{3}\ln \left(\frac{V_f}{V_{th}}\right) ,
\end{equation}
and $V_f$ is the thermodynamic volume at the end of inflation
For the minimal coupled scenario, we have
\begin{equation}
    \label{eq: F(phi) min}
    \begin{split}
    F(\phi) &= \frac{M_{pl}^2}{3} \left[ \frac{d V(\phi)}{d\phi}\right]^{2}\frac{1}{V(\phi)}     ,
    \end{split}
\end{equation}
\begin{equation}
    \Delta N = - \frac{1}{M_{pl}^2} \int_{\phi_i}^{\phi}\frac{V(\chi)}{\partial_{\chi}V(\chi)} d\chi .
\end{equation}
And for the non-minimally coupled case we get
\begin{equation}
    \begin{split}
    F(\phi) &= (1+2\zeta)\frac{\left(2V(\phi)\frac{\partial f}{\partial\phi} - f(\phi) \frac{\partial V}{\partial \phi}\right)^2}{3V(\phi)f^3(\phi)} M_{pl}^2 \\
    & + \frac{14}{3}\zeta\phi \frac{2V(\phi)\frac{\partial f}{\partial\phi} - f(\phi) \frac{\partial V}{\partial \phi}}{f^2(\phi)} \\
    & + \frac{4}{3}\frac{V(\phi)}{f(\phi)} \frac{\zeta \phi^2}{M_{pl}^2} - \frac{\zeta}{3}\phi^2 R(\phi)    ,
    \end{split}
\end{equation}
\begin{equation}
    \begin{split}
    \Delta N &= \int_{\phi_i}^{\phi}\frac{f(\chi)+6\zeta^2\frac{\chi^2}{M_{pl}^2}}{2f(\chi)\frac{\partial f}{\partial\chi} - f^2(\chi)\frac{1}{V(\chi)} \frac{\partial V}{\partial \chi}} d\chi .
    \end{split}
\end{equation}
For $\zeta = 0$, it is easy to see that we recover the functions for the minimally coupled case.

\section{Temperature definitions for the inflaton field} \label{section3}

To fully determine the thermodynamic information for the inflaton field, we require either all the EoS or the fundamental equation.

In the previous section, an EoS for the pressure was derived within the slow-roll approximation. The next step is to determine an EoS for the temperature. In general, there are two approaches: the first relies on the fact that this system behaves as a real gas that can be approximated as an ideal gas, yielding an EoS of the form
\begin{subequations}
    \begin{align}
      &P = - \rho + F. \\
      &T = \frac{\rho V_{th}}{c_1} + l_2,
    \end{align}
\end{subequations}
where $F$ and $l_2$ are functions of the number of e-foldings or the field $\phi$.

The second approach is the one given in \cite{PhysRevD.45.3429,PhysRevD.47.2302}, where the authors generalized the temperature related to the (statistic) energy density and pressure. This temperature scales as $\propto a^{-1} \propto \exp (\Delta N)$, such as in the case of the temperature defined by calculating the energy density via the distribution function for relativistic species in the early universe. We will study both approaches and their implications for the thermodynamics of the system.

\subsection{Real gas fluid-like temperature}

Here we will suppose that the EoS of the system are those of a real gas, which are given by
\begin{subequations}
    \begin{align}
        \frac{1}{T} & = \frac{c_1}{U} = \frac{\partial S}{\partial U} ,\\
        \frac{P}{T} & = \frac{c_2}{V_{th}} = \frac{\partial S}{\partial V_{th}} ,
    \end{align}
\end{subequations}
with $c_1$ being the heat capacity at constant volume and $c_1+c_2$ the heat capacity at constant pressure.
Then for a real gas-like system, we will suppose that the EoS are given by Eqs. (\ref{eq: EoS 1/t real gas}, \ref{eq: EoS p/t real gas}), since their form corresponds to a generalization of the Van der Waals EoS in the weak interactive limit \footnote{The weak interaction limit corresponds to $a \ll 1$, where $a$ is the parameter that mediates the interaction between the particles constituting the Van der Waals fluid.} \footnote{Note also that if $f(U,V)$, $h(U,V)$ are set to zero, we recover the ideal gas EoS}.
\begin{subequations}
    \begin{align}
        \frac{1}{T} & = \frac{c_1}{U} + f(U,V) = \frac{\partial S}{\partial U}, \label{eq: EoS 1/t real gas} \\
        \frac{P}{T} & = \frac{c_2}{V_{th}} + h(U,V_{th}) = \frac{\partial S}{\partial V_{th}}. \label{eq: EoS p/t real gas}
    \end{align}
\end{subequations}
We further suppose that $f,h \ll 1$, that is, that they are small compared to the first term in Eqs. (\ref{eq: EoS 1/t real gas}) and (\ref{eq: EoS p/t real gas}) respectively \footnote{This is the so called weak interactive limit}. These equations can be combined to obtain the usual EoS $P = P(\rho)$, so we have
\begin{equation}
    \label{eq: Pressure real gas}
    P = \frac{c_2}{c_1}\frac{U}{V_{th}} + \frac{U}{c_1}h(U,V_{th}) - \frac{c_2}{c_1^2}\frac{U^2}{V_{th}}f(U,V_{th}) .
\end{equation}
The first term corresponds to the ideal gas, whereas the second and third terms can be set to be a pure function of the volume so that it behaves as the Van der Waals EoS $F(V_{th})$, i.e, we define
\begin{subequations}
    \begin{align}
        F (V_{th}) & =  \frac{U}{c_1}h(U,V_{th}) - \frac{c_2}{c_1^2}\frac{U^2}{V_{th}}f(U,V_{th}) , \\
        h(U,V_{th}) & = \frac{\lambda(V_{th})}{U} , \\
        f(U,V_{th}) & = \frac{\mu (V_{th})}{U^2} .
    \end{align}
\end{subequations}
Bearing this in mind, Eqs.  (\ref{eq: EoS 1/t real gas}) and (\ref{eq: EoS p/t real gas}) become
\begin{subequations}
    \begin{align}
        \frac{1}{T} & = \frac{c_1}{U} + \frac{\mu(V_{th})}{U^2} = \frac{\partial S}{\partial U} ,   \label{eq: sist fund eq 1}\\
        \frac{P}{T} & = \frac{c_2}{V_{th}} + \frac{\lambda(V_{th})}{U} = \frac{\partial S}{\partial V_{th}} . \label{eq: sist fund eq 2}
    \end{align}
\end{subequations}
Now, since the EoS might be consistent with the laws of thermodynamics, a fundamental equation for the system should exist. This requirement implies that the mixed second-order partial derivatives of the fundamental equation should be equal. This condition, in turn, imposes specific restrictions on the functions $\mu(V_{th})$ and $\lambda(V_{th})$,
\begin{equation}
    \label{eq: construction mu lambda}
    \frac{d \mu (V_{th})}{d V_{th}} + \lambda(V_{th}) = 0 .
\end{equation}
For a given function $F(V_{th}) = \lambda(V_{th})/c_1 - (c_2/c_1^2) \mu(V_{th})$, the final condition yields a linear first-order ordinary differential equation for $\mu(V_{th})$. Its solution allows us to determine $\lambda(V_{th})$, and therefore to express the temperature EoS as:
\begin{equation}
    T_{RG} \approx \frac{U}{c_1} + \frac{c_2}{c_1}V_{th}^{-\frac{c_2}{c_1}}\int^{V_{th}} F(\widetilde{V}) \widetilde{V}^{\frac{c_2}{c_1}} d\widetilde{V} .
\end{equation}
Since $\rho = U/V_{th}$, $c_2/c_1 = -1$ and $V_f\exp (-3\Delta N)=V_{th}$, with $V_f$ the thermodynamic volume at the end of inflation,  then the temperature becomes a function of the number of e-foldings
\begin{equation}
    \label{eq: thermodynamic temp eos scale factor}
    T_{RG} = \frac{V_f}{c_1}\exp (-3\Delta N) \left[ \rho(\Delta N) - 3c_1\int^{\Delta N} F(m)dm \right] .
\end{equation}

We identify $c_1$ as a constant that can be set with CMB conditions, and $c_2<0$, implying that the inflaton might have a negative heat capacity, therefore being an unstable system from a thermodynamical point of view; and $F(\Delta N)$, those defined in Sect. \ref{section 2}.

In general, one may use the scalar field itself as the evolution variable in place of the number of e-folds, by directly integrating the field equation of motion \eqref{eq: field evolution}. In many models, however, it is convenient to introduce a normalized or rescaled variable $x$, i.e., typically the inflaton divided by an appropriate energy scale, such as $M_{\rm Pl}$.

\begin{figure}[!htb]
    \includegraphics[width=.49\linewidth]{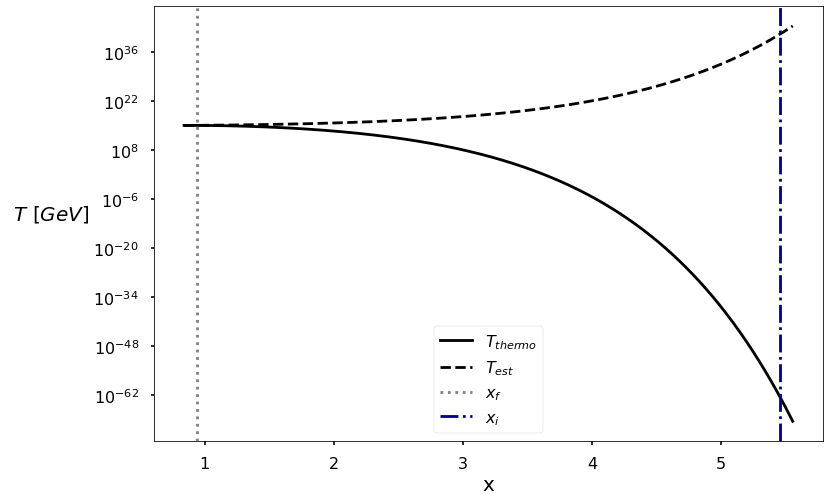} \hfill
    \includegraphics[width=.49\linewidth]{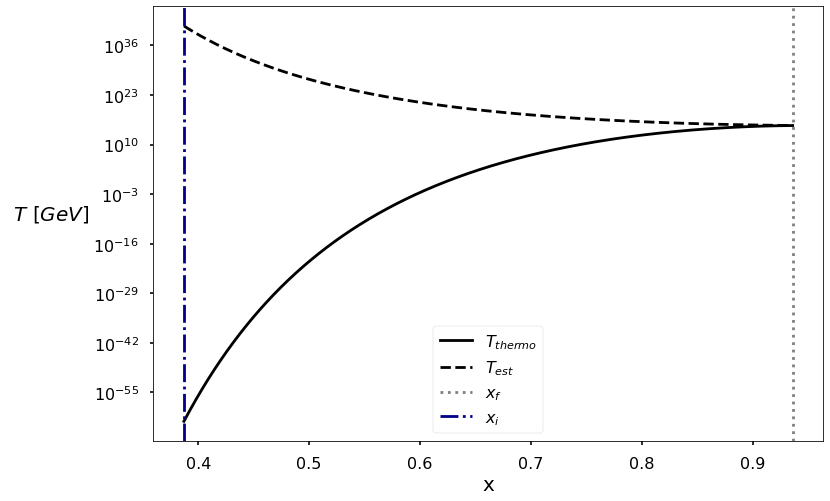} \\
    \vspace{5mm}

    \includegraphics[width=.49\linewidth]{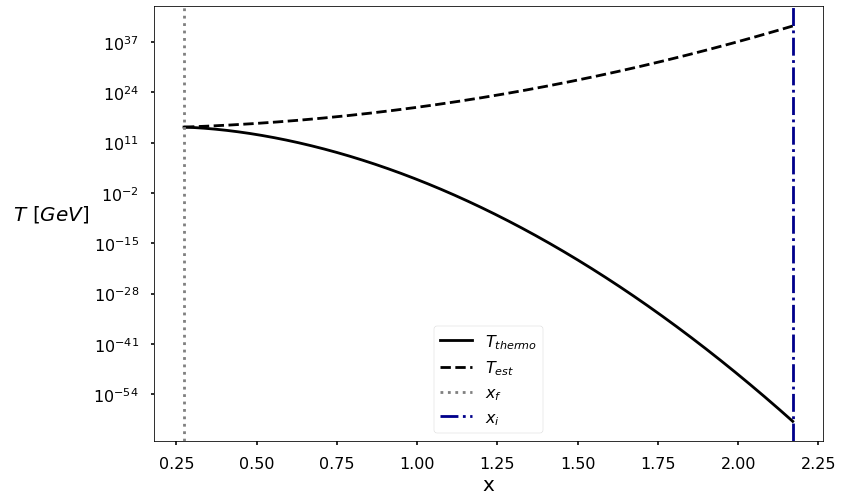}\hfill
    \includegraphics[width=.49\linewidth]{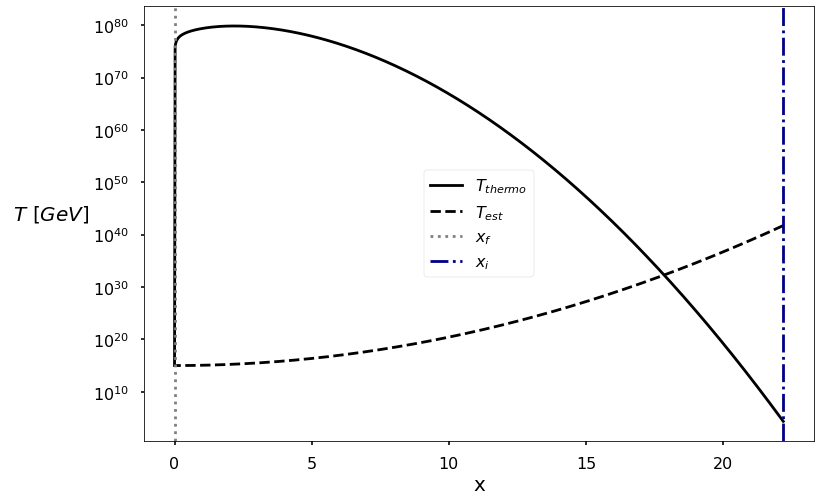}
\caption{Real gas fluid-like temperature (black line) and relativistic species-like temperature (dashed line) as a function of $x=\phi / M_{Pl}$. The temperature is fixed so that at the end of inflation we have $T_{RG}(x_f) = T_{RS}(x_f) = T_f$ . The blue dashed-dotted vertical represents the value of $x$ when inflation started (we suppose $60$ e-foldings), whereas the black dotted vertical line represents the value of $x$ when inflation ends. On the top left Starobinsky, top right hilltop ($n=4$), lower left non-minimally coupling in the Einstein frame, lower right non-minimally coupling in the Jordan frame}
\label{fig: temperature comparison 4 models}
\end{figure}

Bearing this in mind, we can cast the temperature in function of $x$,
\begin{equation}
    \label{eq: thermodynamic temp eos}
    T_{RG}= \frac{V_f}{c_1}\exp (-3\Delta N(x)) \left[ \rho(x) - 3\int^{x} F(x^{\prime}) \frac{d \Delta N}{dx^{\prime}} dx^{\prime} \right].
\end{equation}
For our definitions of $F>0$, $d \Delta N/dx^{\prime} <0$, therefore the sign of the second term in the bracket is positive, implying that the real gas fluid-like temperature is positive, as can be seen in Fig. \ref{fig: temperature comparison 4 models}.

\subsection{Relativistic species-like temperature}

In Ref.   \cite{PhysRevD.45.3429,PhysRevD.47.2302}, it has been argued a first hint to generalize temperature for radiation and scalar field energy density via Eq. (\ref{eq: statistic temp non massive}), having
\begin{equation}
    \label{eq: statistic temp non massive}
    \begin{split}
    \frac{\Dot{T}_{RS}}{T_{RS}} &= - \frac{\Dot{a}}{a}  ,     \\
    T_{RS} &= T_f \exp\left[ \Delta N  \right] .
    \end{split}
\end{equation}
This definition of temperature yields $T_{RS} \propto a^{-1}$, analogous to the scaling in a radiation-dominated universe. We can therefore interpret it as a generalization of the temperature for relativistic species.

The statistical distribution of any particle species depends on the temperature, energy $E$, and chemical potential $\mu$. For relativistic particles, energies are sufficiently high such that the non-degeneracy condition ($\mu/T \ll 1$) and the ultra-relativistic limit ($m/T \ll 1$) hold. This allows us to calculate the energy density via Eq.~\eqref{eq: energy density est}:
\begin{equation}
    \label{eq: energy density est}
    \rho = \int_{0}^{\infty} E \, \mathcal{F} \, d^3\Vec{p},
\end{equation}
where $\mathcal{F}$ denotes the distribution function, i.e., either fermions or bosons, and $\Vec{p}$ the momentum.

Given the conditions above, the integrand depends only on the magnitude of the momentum $l \equiv |\Vec{p}|$, since for ultra-relativistic particles, the dispersion relation simplifies to $E \approx l$.

Accounting for the degeneracy of relativistic species, the energy density acquires the usual form,
\begin{equation}
    \label{eq: rho_rad_temp}
    \rho(T_{RS}) = g_{\ast} \frac{\pi^2}{30} T_{RS}^4,
\end{equation}
where $g_{\ast}$ denotes the effective number of relativistic degrees of freedom.

For a relativistic fluid, the energy density scales as $\rho \propto a^{-4}$.

Consequently, from Eq. (\ref{eq: rho_rad_temp}), the temperature scales as $T_{RS} \propto a^{-1}$, i.e., consistent with our initial definition.

\section{Fundamental Equation for $T_{RG}$}

From Eqs. (\ref{eq: sist fund eq 1}) and (\ref{eq: sist fund eq 2}), we can obtain the fundamental equation related to $T_{RG}$. Hence, writing
\begin{equation}
    \begin{split}
    \label{eq: real gas fun equation}
    S_{RG} (U,V_{th}) =& c_1 \ln\left( U\right) + c_2 \ln \left( V_{th} \right) \\
    &+ \frac{c_1}{U} V_{th}^{-\frac{c_2}{c_1}}\int^{V_{th}} F \left( \widetilde{V}\right) {\widetilde{V}}^{\frac{c_2}{c_1}} d \widetilde{V},
    \end{split}
\end{equation}
and setting $c_2/c_1 = -1$, we end up with
\begin{equation}
    \label{eq: thermodynamic fundamental equation}
    S_{RG} (x) = c_1 \left[\ln\left( \rho(x) \right) - \frac{3}{\rho(x)}\int^{x} F(x^{\prime}) \frac{d \Delta N}{dx^{\prime}} dx^{\prime} \right] .
\end{equation}
The above entropy function is well defined. It turns out to be positive and to increase with respect to $U$ and $V_{th}$. Additionally, it also goes to a minimum whenever the temperature tends to zero. As expected, therefore, Eq. (\ref{eq: real gas fun equation}) appears as a fundamental equation. Hence, the laws of thermodynamics may be naturally addressed through it, enabling us to explore possible phase transitions and critical points during the overall inflationary epoch.

\subsection{Isotherms}

From the fundamental equation, we can calculate the isotherms of this system, inverting the temperature EoS, i.e. calculating $U = U(T_{RG})$, and performing the variable change in the pressure EoS (\ref{eq: Pressure real gas})
\begin{widetext}
\begin{subequations}
    \begin{align}
        P (V_{th},T_{RG}) &=T_{RG} \left \{ \frac{c_2}{V_{th}} + \frac{c_1}{U(V_{th}, T_{RG})} F(V_{th}) - \frac{c_2 }{U(V_{th},T_{RG})} V_{th}^{-\frac{c_2}{c_1}-1}\int^{V_{th}} F \left( \widetilde{V}\right) {\widetilde{V}}^{\frac{c_2}{c_1}} d \widetilde{V} \right\} ,
     \\
        U  (V_{th}, T_{RG}) = & \, c_1T_{RG} -  \frac{1}{c_1}V_{th}^{-\frac{c_2}{c_1}}\int^{V_{th}} F \left( \widetilde{V}\right) {\widetilde{V}}^{\frac{c_2}{c_1}} d \widetilde{V}    .
    \end{align}
\end{subequations}
Then again, setting $c_2/c_1 = -1$
\begin{subequations}
    \begin{align}
        P (V_{th}, T_{RG}) & = T_{RG} \left\{- \frac{c_1}{V_{th}} + \frac{c_1}{U(V_{th}, T_{th})} F(V_{th}) +  \frac{c_1}{U(V_{th},T_{RG})} \int^{V_{th}} F \left( \widetilde{V}\right) {\widetilde{V}}^{-1} d \widetilde{V} \right\}   ,
        \label{eq: thermodynamic isoterm} \\
        U(V_{th}, T_{RG}) & = c_1T_{RG} -  \frac{1}{c_1}V_{th}\int^{V_{th}} F \left( \widetilde{V}\right) {\widetilde{V}}^{-1} d \widetilde{V} .     \label{eq: U(V, T_{thermo})}
    \end{align}
\end{subequations}
\end{widetext}

If the isotherm has a boundary where its behavior changes, such as becoming an increasing function of volume in one region while remaining a decreasing function of volume elsewhere, this may indicate the presence of a phase transition.

However, the \emph{stability conditions} enable us to determine algebraically the point at which a phase transition occurs.

\subsection{Stability Conditions}

The stability conditions are given by
\begin{subequations}
    \begin{align}
    \frac{\partial^2 S}{\partial U^2} & \leq 0 ,  \label{eq: global stability condition 1} \\
    \frac{\partial^2 S}{\partial V_{th}^2} & \leq 0  ,\label{eq: global stability condition 2} \\
    \frac{\partial^2 S}{\partial U^2}\frac{\partial^2 S}{\partial V_{th}^2} - \left(\frac{\partial^2 S}{\partial U\partial V_{th}}\right)^2  &  \geq 0 . \label{eq: local stability condition}
    \end{align}
\end{subequations}

If the system fails to satisfy these conditions at some $U$ and $V$, a phase transition occurs at that point. For the real gas fundamental equation, these conditions are expressed as follows
\begin{widetext}
\begin{subequations}
    \begin{align}
    -\frac{c_1}{U^2} + \frac{2c_1}{U^3}\int F(\widetilde{V})\widetilde{V}^{-1} d\widetilde{V}& \leq 0  , \label{eq: global stability condition 1 thermo} \\
    -\frac{c_1}{V_{th}^2} + \frac{c_1}{U}\frac{F^{\prime}(V_{th})}{V_{th}} - \frac{c_1}{U}\frac{F(V_{th})}{V_{th}^2}& \leq 0 , \label{eq: global stability condition 2 thermo} \\
    \left[ -\frac{c_1}{U^2} + \frac{2c_1}{U^3}\int F(\widetilde{V})\widetilde{V}^{-1} d\widetilde{V}\right] \left[-\frac{c_1}{V_{th}^2} + \frac{c_1}{U}\frac{F^{\prime}(V_{th})}{V_{th}} - \frac{c_1}{U}\frac{F(V_{th})}{V_{th}^2}\right] - \frac{c_1^2}{U^4}\frac{F^2(V_{th})}{V_{th}^2}&  \geq 0 .\label{eq: local stability condition thermo}
    \end{align}
\end{subequations}
\end{widetext}

\subsection{Critical Points}

The existence of critical points is determined by the following conditions:
\begin{equation}
    \label{eq: critical point conditions}
    \left.\frac{\partial P}{ \partial V}\right|_{T_{cr},V_{cr}} = \left.\frac{\partial^2 P}{ \partial V^2}\right|_{T_{cr},V_{cr}} = 0 .
\end{equation}
These two conditions will give us a system of algebraic equations for $V_{cr}$ and $T_{cr}$. For the real gas temperature, the critical point conditions are
\begin{widetext}
\begin{subequations}
    \begin{align}
    \label{eq: crit point cond 1 thermo}
    &\left. c_1 \left[\frac{F^{\prime}(V_{th})}{U} + \frac{F(V_{th})}{UV_{th}} + \frac{1}{V_{th}^2}\right] + \frac{1}{U^2}\left[F(V_{th}) + \int^{V_{th}} F(\widetilde{V}) \widetilde{V}^{-1} d\widetilde{V}\right]^2\right|_{T_{cr};V_{cr}} = 0,\\
    &\left.c_1 \left[\frac{F^{\prime \prime}(V_{th})}{U^2} - \frac{F^{\prime}(V_{th})}{UV_{th}} - \frac{F(V_{th})}{UV_{th}^2} - \frac{2}{V_{th}^3}\right] + \frac{2}{c_1U^3}\left[F(V_{th}) + \int^{V_{th}} F(\widetilde{V}) \widetilde{V}^{-1} d\widetilde{V}\right]^3 \right. \notag\\
    &\left. + \frac{3}{U^2}\left[ \frac{F^3(V_{th})}{V_{th}} + F(V_{th})F^{\prime}(V_{th}) + \frac{F(V_{th})}{V_{th}}\int^{V_{th}} F(\widetilde{V}) \widetilde{V}^{-1} d\widetilde{V} + F^{\prime}(V_{th})\int^{V_{th}} F(\widetilde{V}) \widetilde{V}^{-1} d\widetilde{V} \right]\right|_{T_{cr};V_{cr}} = 0 .\label{eq: crit point cond 2 thermo}
    \end{align}
\end{subequations}
\end{widetext}

\section{Fundamental Equation for $T_{RS}$}

For the pressure and temperature, Eqs. (\ref{eq: Pressure EoS}) and (\ref{eq: statistic temp non massive}), respectively, there are no EoS under the formal forms $P/T_{RS} = P/T_{RS} (U, V_{th})$, $1/T_{RS} = 1/T_{RS} (U, V_{th})$ and, so, we can proceed by supposing a structure $P/T_{RS}$ as
\begin{equation}
    \label{eq: ansatz statistic pressure eos}
    \frac{P}{T_{RS}} = - \frac{U + V_{th}F(V_{th})}{T_fV_{th}^{2/3}V_f^{1/3}} .
\end{equation}
Since $\partial S_{RS}/ \partial V_{th} = P/T_{RS}$ for $S_{RS} = S_{RS}(U,V_{th})$, we can write
\begin{equation}
    \label{eq: ansatz statistic fund equation}
    S_{RS}(U,V_{th}) = - \int \frac{U + V_{th}F(V_{th})}{T_fV^{2/3}V_f^{1/3}} dV  + F_2(U),
\end{equation}
for some $F_2(U)$. Now, since $S_{RS}$ is a fundamental equation, then $\partial S_{RS}/ \partial U = 1/T_{RS}$, therefore, we obtain
\begin{equation}
    \label{eq: ref 1}
    \frac{d F_2(U)}{dU} =  \frac{4}{T_{RS}}.
\end{equation}

Given that this EoS should reduce to $T_{RS}$, as in Eq. (\ref{eq: statistic temp non massive}), we may propose an ansatz for the dependence of the internal energy $U$ on the scale factor.

A natural choice may arise from the fact that, during inflation, $\rho \approx cte$, and since $V_{th} \propto \exp (-3\Delta N)$, it follows that $U \propto \exp (-3\Delta N)$.

Substituting $U = U_f \exp (-3\Delta N) $, having $U_f$ the value of the internal energy at the end of inflation, into Eq.  (\ref{eq: ansatz statistic fund equation}), and using the relation (\ref{eq: ref 1}), we ultimately obtain $F_2$ as
\begin{equation}
    \label{eq: H(U)}
    F_2(U) =  \frac{3}{T_f}\frac{U^{4/3}}{\rho_f^{1/3}V_f^{1/3}}.
\end{equation}
Hence, the fundamental equation for this system becomes
\begin{align}
    \label{eq: statistic fund equation}
    &S_{RS}(U,V_{th}) =\\ - &\frac{1}{T_fV_f^{1/3}}\left[3UV_{th}^{1/3} + \int^{V_{th}}\widetilde{V}^{1/3}F(\widetilde{V}) d\widetilde{V}\right] +\frac{3}{T_f}\frac{U^{4/3}}{\rho_f^{1/3}V_f^{1/3}},
\end{align}
whereas the temperature reads
\begin{equation}
    T_{RS}(U,V_{th}) = \frac{T_f}{3}\left(\frac{V_f}{V_{th}}\right)^{1/3}\left[\frac{4}{3}\left(\frac{U}{\rho_f V_{th}}\right)^{1/3} -1 \right]^{-1} .
\end{equation}
Here, $\rho_f$ and $V_f$ represent the values of the energy density and volume at the end of inflation, respectively. In terms of $x$, the fundamental equation takes the form
\begin{equation}
    \label{eq: statistic fund equation func de x}
    S_{RS}(x) = - \frac{3V_f}{T_f}\int^{x} F(x^{\prime})\exp \left(-4\Delta N (x^{\prime}) \right)\frac{d \Delta N}{dx^{\prime}} dx^{\prime} .
\end{equation}
Here, again since $d \Delta N/dx^{\prime}<0$, the whole expression, Eq. (\ref{eq: statistic fund equation func de x}), turns out to be positive.

It is important to note that this may not be the only fundamental equation whose EoS reduces to the desired ones. Clearly, under our assumptions, it appears particularly simple though.

Nevertheless, Eq. (\ref{eq: statistic fund equation})  possesses several advantageous features, as listed below.

\begin{itemize}
    \item[-] It represents a continuously differentiable function of $U$ and $V$, maximal with respect to these variables, and monotonically increasing with respect to the internal energy.
    \item[-] It additionally matches the third law of thermodynamics, as the entropy approaches a minimum when the temperature vanishes.
\end{itemize}

We can now investigate the thermodynamical behavior of this definition of temperature.

\subsection{Isotherms}

We can follow the same procedure as in the $T_{RG}$ isotherms to calculate the isotherms for $T_{RS}$,
\begin{widetext}
\begin{subequations}
    \begin{align}
    P (V_{th}, T_{RS}) & = -\frac{T_{RS}}{T_fa_fV_0^{1/3}}\frac{1}{V_{th}^{2/3}} \left[U(V_{th},T_{RS})+ V_{th}F(V_{th}) \right] , \label{eq: statistic isoterm} \\
    U(V_{th}, T_{RS}) & = \frac{\rho_f}{64}\left[V_0^{1/3}a_f\left(\frac{T_f}{T}\right) + 3V_{th}^{1/3}\right]^3 . \label{eq: U(V, T_{est})}
    \end{align}
\end{subequations}
\end{widetext}

\subsection{Stability Conditions}

For $T_{RS}$, we have the stability conditions
\begin{widetext}
\begin{subequations}
    \begin{align}
    \frac{4}{3}\frac{1}{T_f\rho_f^{1/3}V_f^{1/3}}\frac{1}{U^{2/3}} & \leq 0 ,  \label{eq: global stability condition 1 est} \\
    -\frac{1}{T_fV_f^{1/3}}\left[-\frac{2}{3}\frac{U}{V_{th}^{5/3}} + \frac{1}{3}\frac{1}{V_{th}^{2/3}}F(V_{th}) + V_{th}^{1/3} F^{\prime}(V_{th}) \right]& \leq 0 , \label{eq: global stability condition 2 est} \\
    -\left(\frac{1}{T_fV_f^{1/3}}\right)^2\left\{ \frac{4}{3\rho_f^{1/3}}\frac{1}{U^{2/3}}\left[-\frac{2}{3}UV_{th}^{-5/3} + \frac{1}{3}V_{th}^{-2/3}F(V_{th}) + V_{th}^{1/3}F^{\prime}(V_{th})\right] + \frac{1}{V_{th}^{4/3}} \right\}&  \geq 0 . \label{eq: local stability condition est}
    \end{align}
\end{subequations}
\end{widetext}
From this, it is clear that a phase transition occurs only if all the conditions,  functions of the scale factor or of
$\phi$, simultaneously fail at the same scale factor or $\phi$ value.

\subsection{Critical Points}

Conditions (\ref{eq: critical point conditions}) provide a system of algebraic equations for the critical values of the extensive variables, $V_{cr}$ and $T_{RS,cr}$, as reported below.
\begin{subequations}
\begin{align}
      0 = & -\frac{2}{3} \frac{U}{V_{th}^{5/3}} + \frac{\partial U}{\partial V_{th}}\frac{1}{V_{th}^{2/3}} + \label{eq: crit point cond 1 est} \\
      & +\frac{1}{3}\frac{F(V_{th})}{V_{th}^{2/3}} + \,  F^{\prime}\left.(V_{th})V_{th}^{1/3} \right|_{T_{cr};V_{cr}}. \notag \\
      0 = &\frac{10}{9}\frac{U}{V_{th}^{8/3}} - \frac{2}{3}\frac{\partial U}{\partial V_{th}}\frac{1}{V_{th}^{5/3}} + \frac{1}{V_{th}^{2/3}}\frac{\partial^2 U}{\partial V_{th}^2} - \label{eq: crit point cond 2 est} \\
    &- \frac{4}{9}\frac{F(V_{th})}{V_{th}^{5/3}} \left. + \frac{4}{3}\frac{F^{\prime}(V_{th})}{V_{th}^{2/3}} + F^{\prime\prime}(V_{th})V_{th}^{1/3} \right|_{T_{cr};V_{cr}}  .  \notag
\end{align}
\end{subequations}

\section{Physical results} \label{section4}

We analyzed \emph{six different potentials} using for them the two temperature definitions, along with their respective fundamental equations, stability conditions, and critical points.

We selected those models since they turn out to be the most viable inflationary paradigms as reported in Ref. \cite{Plank2018inflation}.

Nevertheless, we selected the statistically most suitable bounds over their parameters, as provided in Ref. \cite{encyclopedia}.

For the sake of simplicity, each scenario, along with each potential, is dubbed in a concise way, recalling their names, as reported in what follows.

\subsubsection{Large field inflationary models}

These models are characterized by a super-Planckian field excursion, $\Delta \phi > M_{pl}$, during the inflationary epoch, where $\Delta \phi$ is the total field variation from the time observable CMB scales, exit the horizon until the end of inflation. These models exhibit several distinctive features: a red-tilted scalar spectrum ($n_s < 1$) consistent with Planck data, detectable primordial tensor modes that produce a significant tensor-to-scalar ratio ($r \gtrsim 0.01$), and a smooth, monotonic potential that typically satisfies $V'(\phi) > 0$ throughout inflation.

\begin{itemize}
    \item[-]
    Starobinsky in the Einstein frame. \cite{STAROBINSKY198099,DAVIES1977108,PhysRevD.32.2511, PhysRevD.39.3159,De_Felice_2010}
    \begin{equation}
         \label{eq: starobinsky potential}
         V(\phi) = M^4\left[1-\exp\left(-\sqrt{\frac{2}{3}}\frac{\phi}{M_{pl}}\right)\right]^2.
     \end{equation}
    \item[-]
    E-model for $n=1,2$ \cite{Kallosh_2013_2} .
    \begin{equation}
        \label{eq: Emodel potential}
        V(\phi) = M^4\left[1-\exp\left(-\sqrt{\frac{2}{3\alpha}}\frac{\phi}{M_{pl}}\right)\right]^{2n}.
    \end{equation}
    \item[-]
    T-model with $n=1,2$\cite{Kallosh_2013,Kallosh_2013_2}
    \begin{equation}
        \label{eq: Tmodel potential}
        V(\phi) = M^4 \left[ \tanh \left(\frac{1}{\sqrt{6\alpha}}\frac{\phi}{M_{pl}}\right) \right]^{2n}.
    \end{equation}
    \item[-]
    Non-minimally coupling with a Higgs-like potential in both the Einstein and Jordan frames \cite{BEZRUKOV2008703,BEZRUKOV200988,Bezrukov_2009,PhysRevD.84.123504}.
    \begin{equation}
        \label{eq: Higgs-like potential}
        V(\phi) = M^4\left(\frac{\phi^2}{M_{pl}^2}-\frac{v^2}{M_{pl}^2}\right)^2.
    \end{equation}
\end{itemize}

\subsubsection{Small field inflationary models}

These models are characterized by a sub-Planckian field excursion, $\Delta \phi < M_{\text{pl}}$, during the inflationary epoch, where $\Delta \phi$ is the total field variation from the time observable CMB scales, exit the horizon until the end of inflation. Their key features include a red-tilted scalar spectrum ($n_s < 1$), consistent with Planck data, and a small tensor-to-scalar ratio ($r \lesssim 0.01$) implying negligible primordial gravitational waves. From a theoretical perspective, these potentials are more readily embedded in high-energy theories like string theory and supergravity, which allows for a natural \emph{ultraviolet completion}.

\begin{itemize}
    \item[-] hilltop minimally and non-minimally coupled for $n=2,4$ \cite{LINDE1982389,LINDE1983317,PhysRevD.75.123510,Boubekeur_2005}
    \begin{equation}
        \label{eq: hilltop potential}
        V(\phi) = M^4\left[1-\left(\frac{\phi}{\mu}\right)^n\right].
    \end{equation}
    \item[-] hybrid inflation, along the valley of one field, i.e. the field set to zero, in minimally and non-minimally coupled, say \cite{PhysRevD.49.748,LYTH19991,covi2000modelsinflationsupersymmetrybreaking,lazarides2001supersymmetrichybridinflation,panagiotakopoulos2000hybridinflationsupergravity}
    \begin{equation}
        \label{eq: vhi potential}
        V(\phi) = M^4\left[1+ \left(\frac{\phi}{\mu}\right)^2 \right].
    \end{equation}
\end{itemize}

\subsubsection{Natural inflation}
Natural inflation \cite{Kim_2005,PhysRevD.47.426,PhysRevLett.65.3233,PRESKILL1983127,LINDE1988437} employs a pseudo-Nambu-Goldstone boson (axion) as the inflaton, where the flatness of the potential is protected by a shift symmetry $\phi \to \phi + \text{constant}$. This symmetry is broken non-perturbatively, generating the periodic potential given in Eq. (ref{eq: natural inflation potential}). The behavior of the model depends crucially on the decay constant $f$. For $f \gtrsim M_{pl}$, it behaves as a large-field model ($\Delta \phi > M_{pl}$), whereas for $f \ll M_{pl}$, it becomes a small-field model ($\Delta \phi < M_{pl}$). However, to match observations of CMB scales, the field excursion should satisfy $\Delta \phi \approx \pi f$, which imposes the constraint $f > M_{pl}/\pi$.

We intend to study this case, focusing on both minimally and non-minimally coupled scenarios:
\begin{equation}
        \label{eq: natural inflation potential}
        V(\phi) = M^4\left[1+\cos \left(\frac{\phi}{f}\right) \right].
    \end{equation}

For the non-minimally coupled models, thermodynamics was studied in both the Einstein and Jordan frames. Here, we summarize our findings.

First, $T_{RG}$ tends to increase as we approach the end of inflation for all minimally coupled models and the non-minimally coupled models in the Einstein frame. This behavior is counterintuitive to the conventional understanding of inflation. In the Jordan frame, a similar trend is observed. In contrast, $T_{RS}$ decreases as we approach the end of inflation for all models and in both frames, as illustrated in Figure \ref{fig: temperature comparison 4 models} in Appendix \ref{appendix plots}.

The entropy for minimally coupled models and in the Einstein frame is always well-defined and follows the third law of thermodynamics for both temperatures; it decreases when the temperature also decreases. So in principle, both definitions of temperature can fulfill the laws of thermodynamics.

$T_{RS}$ allows the system to have phase transitions and critical points; as shown in the figure \ref{fig: isotherm comparison 2 models} in Appendix \ref{appendix plots}, there is a portion of the isotherms that mimics the Van der Waals isotherms, meaning that a region may exist where two phases coexist or, depending on the conditions, achieve a critical point.

Hence, the relativistic species-like temperature, $T_{RS}$, represents the physical temperature of the inflaton field. Since the inflaton field follows the laws of thermodynamics, $T_{RS}$ would be the field's observable temperature during the inflationary epoch. In other words, it is the temperature that we would be able to measure.

\subsection{Minimally Coupled Models}

The \emph{$\alpha$-attractor models}, such as the T- and E-models in the Jordan frame, are constructed from a non-canonical kinetic term for the inflaton field, placing their fundamental formulation beyond the immediate scope of our formalism. However, their corresponding potentials in the Einstein frame are well-defined. Consequently, we will treat them as minimally coupled models within this analysis, as we study them exclusively in the Einstein frame and their Lagrangian does not feature an explicit coupling to gravity.

\subsubsection{Outputs for large fields}

For the \emph{Starobinsky, E-model, and T-model} potentials with $T_{RS}$, the second derivative of the entropy with respect to volume is positive throughout inflation, violating a key stability condition. This behavior is characteristic of a second-order phase transition. Since the entire inflationary epoch is metastable, a perturbation would trigger a phase transition. We identify the onset of this metastability—the beginning of inflation, which occurs for $\Delta N > 60$—as the moment the phase transition occurs. Consequently, we interpret this phase transition as the mechanism that initiated inflation.

The equations defining the critical points do not reach zero. Accordingly, no critical point exists for these potentials.

It is not surprising that all these scenarios exhibit similar thermodynamic behavior, given their dynamical similarities. For example, at the end of inflation, all these potentials can be approximated as $\propto \phi^2$. Furthermore, at the onset of inflation, any T-model can be rewritten in the form $\propto 1 - 2n\exp(-\sqrt{2/3}\, \widetilde{\phi})$ by redefining $\phi = \sqrt{\alpha/3}\,\widetilde{\phi}$. This is, in fact, the functional form of the Starobinsky and E-model potentials for $\phi \gg 1$.

\subsubsection{Outputs for small fields}

\emph{Hilltop potentials} for $T_{RS}$ exhibit a metastable state at large values of the number of e-foldings ($\Delta N \geq 78$) for both $n=2$ and $n=4$, deep in the Planck epoch. The issue is that classical thermodynamics may not be a well-defined theory in this regime; therefore, we cannot definitively conclude that this model features a phase transition. At these values of $\Delta N$, quantum corrections are required to correctly describe the field's dynamics. On the other hand, the model exhibits a critical point near the end of inflation. This implies that the inflaton behaves as a \emph{supercritical fluid} \cite{doi:10.1021/cr980085a,supercritical_fluid_1}, exhibiting properties of both phases simultaneously. This is equivalent to stating that all symmetries would remain unbroken after inflation (since the phase transition occurred at Planckian times), a scenario which would, of course, require an external mechanism to break them and produce the observable universe of today.

As noted previously, quantum corrections are necessary for this model as it represents an unstable potential  \cite{Kallosh_2019,Kazunorikohri_2010}. If we assume that quantum corrections effectively contribute to a 1-loop self-coupling correction given by a Coleman-Weinberg potential\cite{PhysRevD.7.1888,han2025higgspoleinflationloop} term, $\propto \phi^4 \log(\phi/M_{pl})$. The effective potential is
\begin{equation}
    V_{\text{eff}} \approx M^4 \left[ 1 - a \left( \frac{\phi}{M_{\text{pl}}} \right)^n - b \left( \frac{\phi}{M_{\text{pl}}} \right)^4 \ln \left( \frac{\phi}{M_{\text{pl}}} \right) \right],
\end{equation}
where $a$ and $b$ are positive constants. Assuming again that our thermodynamic formulation remains valid for such potentials and that the constants values $a$, $b$ are such that the dominating term is the Coleman-Weinberg, we arrive at a scenario similar to that of large-field models, where the second derivative of the entropy with respect to volume is positive throughout inflation, implying a second-order phase transition, and no critical points are present.

In summary, quantum corrections qualitatively change the thermodynamic picture, causing the critical point to vanish.

\emph{Hybrid inflation} along the valley of one field (i.e., the field set to zero) was studied in the small-field regime, where $\mu/M_{\rm pl} < 1/\sqrt{2}$ \cite{PhysRevD.80.123534,PhysRevD.79.103507}. In this regime, neither temperature exhibited phase transitions or critical points.

It is interesting to note that at the critical value of $\mu/M_{\rm pl} = 1/\sqrt{2}$, $T_{RS}$  has a critical point near the beginning of inflation. However, this specific value for $\mu/M_{\rm pl}$ is the minimum required for slow-roll inflation; for smaller values, fast-roll occurs, causing the approximations used in this thermodynamic theory to break down. This critical point implies a \emph{supercritical fluid-like} phase, similar to the hilltop potential, leading to an inflationary epoch that resembles the effective behavior of the Planckian primordial universe, a state that would require an external mechanism to transition into the observable universe of today.

\subsubsection{Outputs for natural inflation}

For natural inflation, the second and third stability conditions are not fulfilled for $T_{RS}$ across a wide interval of $\Delta N$. This implies the existence of a metastable region (spanning from $71.55$ to $33.51$ e-foldings for $f = 17M_{pl}$). Therefore, a phase transition might occur at the beginning of inflation ($\Delta N \sim 60$) or even earlier. In this scenario, the phase transition can be interpreted as the trigger of inflation.

No critical points exist for this temperature. It is important to note that this behavior is tied to the value of the decay constant $f$. For $f \gtrsim M_{pl}$, the description given above holds. However, as $f$ decreases, the metastable region shrinks.

\subsection{Non-Minimally Coupled Models}
\label{subsec:nonminimally_coupled}

A priori, for all models we assumed a value of $\zeta = 10^{-4}$ to ensure that the overall sign in the Lagrangian remains unchanged. While this condition is strictly necessary only for large-field models, we maintain a fixed $\zeta$ across all models for consistency. Subsequently, we investigated the thermodynamic consequences of varying the coupling parameter over a range of values.

\subsubsection{Outputs for  large fields}

The \emph{non-minimally coupled  Higgs-like} potential does not exhibit a critical point in both the Einstein and Jordan frames. It is true, however, that the conditions are fulfilled separately, i.e. they become zero at some point of inflation, but they do not coincide.

In parallel, the stability conditions break down in both frames, indicating that a phase transition should occur. In the Jordan frame, two stability conditions (the first and third) are violated near the beginning of inflation. This suggests that the system is in a metastable state for the corresponding field values, implying that a perturbation could trigger a phase transition. In the Einstein frame, a similar metastability \footnote{In this case, it is the second stability condition that is violated.} occurs, also around the \emph{beginning of inflation}. This implies that the phase transition itself might initiate the inflationary epoch. Since physical observations are carried out in the Einstein frame, we can interpret \emph{inflation as a physical process that was initiated by a thermodynamic phase transition}.

Finally, it is important to note that the thermodynamics for this temperature are frame-independent, as both critical points and indications of phase transitions are observed in both frames, and within similar intervals of $\Delta N$.

\subsubsection{Outputs for  small fields}

\emph{Hilltop potentials} for $T_{RS}$ in the Jordan frame do not exhibit a critical point, in contrast to their minimally coupled counterparts. However, the first and third stability conditions are violated for both $n=2$ and $n=4$. We interpret the maximum value of $\Delta N > 60$ at which both conditions are violated as the moment at which a phase transition occurs. This allows us to identify the phase transition as the event that initiated the inflationary epoch.

In the Einstein frame, the potentials for $n=2$ and $n=4$ are, respectively:
\begin{equation}
    \begin{split}
        V(\chi) = M^4\left[1 - \left(\frac{1}{\mu^2} - \frac{2\zeta}{M_{pl}^2}\right)\chi^2 - \frac{2\zeta}{\mu^2 M_{pl}^2}\chi^4 \right] .
    \end{split}
\end{equation}
\begin{equation}
    \begin{split}
        V(\chi) = M^4\left[1 + \frac{2\zeta}{M_{pl}^2}\chi^2 +  \left(+ \frac{2\zeta^2}{M_{pl}^4} - \frac{1}{\mu^4} \right)\chi^4  \right] .
    \end{split}
\end{equation}
These potentials also lack critical points and fail the first and third stability conditions in the same manner. Therefore, the thermodynamic behavior---specifically, the absence of a critical point and the violation of stability conditions indicating a phase transition at the onset of inflation---is \emph{frame-independent}. Furthermore, unlike the minimally coupled case, this non-minimally coupled scenario does not require the introduction of quantum corrections to generate additional terms in the potential.

In \emph{hybrid inflation} along the valley of one field (i.e., with the other field set to zero) within the Jordan frame, the evolution is smooth for energy scales $\mu/M_{pl} < 1/\sqrt{2}$, exhibiting no critical points or phase transitions.
The situation changes dramatically at $\mu/M_{pl} = 1/\sqrt{2}$. Here, the second stability condition for $T_{RS}$ is violated, implying that the system enters a metastable state. We interpret this metastability as the signal of a phase transition that marks the beginning of inflation. This is a key difference from the minimally coupled scenario, where a critical point occurs near the beginning of inflation.

In the Einstein frame, the potential is
\begin{equation}
    V(\chi) = M^4\left[1-\left(\frac{\chi}{\mu}\right)^n\right] \left[1 + 2\zeta\left(\frac{\chi}{M_{pl}}\right)^2\right] .
\end{equation}
However, the thermodynamic description remains unchanged between the frames.

\subsubsection{Outputs for natural inflation}

The potential in the Einstein frame is
\begin{align}
        V(\chi) &=M^4\left[1+\cos \left(\frac{\chi}{f}\right) + 2\zeta\left(\frac{\chi}{M_{pl}}\right)^2\right.\\
        &\left.+2\zeta\left(\frac{\chi}{M_{pl}}\right)^2\cos \left(\frac{\chi}{f}\right) \right],
\end{align}
and has no critical points or phase transitions, as all stability conditions are fulfilled.

This contrasts with the minimally coupled scenario, where a (periodic) phase transition occurs. The difference in thermodynamic behavior is likely due to the strong dependence of natural inflation on the coupling parameter $\zeta$. While values of $\zeta > 10^{-3}$ are observationally preferred \cite{Bostan:2022swq}, we have fixed $\zeta = 10^{-4}$ for consistency with the other models in our analysis. Below, we  discuss the impact of having larger values of $\zeta$ on our outputs, concluding that very small values appear to better agree with our theoretical scheme.

\subsubsection{Selecting small $\zeta$ and bounds on non-minimal couplings}

Several models of Higgs inflation show that the coupling constants may be taken large \cite{Nozari:2024hip,Branchina:2025hen}, albeit in contrast to other works, where the couplings might be small; instead, see Ref. \cite{Kaganovich:2025iht}, or the limits found by the Planck collaboration \cite{Plank2018inflation}.

Here, we further explore the thermodynamic behavior of our frameworks for different $\zeta$. For $\zeta < 1/6$, the thermodynamic picture for large-field models remains largely unchanged, as the instability indicated by the second stability condition persists in both frames. However, at $\zeta = 1/6$—the value that enables a conformal mapping of the action in Eq. (\ref{eq: accion campo escalar nonmin})—the thermodynamic behavior shifts. At this critical coupling, both the second and third stability conditions indicate thermodynamic instabilities, suggesting the occurrence of a phase transition rather than a metastable phase.

For $\zeta > 1/6$, the sign in the action (\ref{eq: accion campo escalar nonmin}) changes, and in the Einstein frame, the potential turns out to be
\begin{equation}
V_E(\phi) = \frac{V_J(\phi)}{(1 - \xi \phi^2 / M_{pl}^2)^2},
\end{equation}
and may exhibit divergences if, for example, it holds that $\zeta = 1$,  $\phi = M_{pl}$.

For large-field models, these divergences occur at progressively larger field values as $\zeta$ increases, ultimately breaking the frame-independent thermodynamic behavior.

Conversely, small-field models can accommodate larger values of $\zeta$ without necessarily encountering these divergences within the inflationary field range. For the hilltop potential at $\zeta = 1/6$, only the second stability condition is violated, signaling a transition from a possible first-order phase transition, i.e., for $\zeta < 1/6$, to a second-order phase transition\footnote{We recall that the difference between first and second order phase transitions consists of the fact that first-order transitions proceed via bubble nucleation and are characterized by a discontinuous jump of the order parameter, whereas second-order transitions occur smoothly, with the order parameter vanishing continuously as the system approaches criticality and the correlation length diverges, see Ref. \cite{Quiros:1999jp}.}.

At larger values, such as $\zeta \geq 1$, a critical point emerges near the beginning of inflation, reintroducing the problematic supercritical fluid behavior, observed in the minimally coupled case.

For hybrid inflation along the valley of one field, the scenario exhibits metastability with the critical value $\mu/M_{pl} = 1/\sqrt{2}$ for $\zeta \leq 1/6$. At higher $\zeta$ values, the potential in the Einstein frame effectively becomes quadratic and exhibits phase transitions near the inflationary onset, albeit this occurrence \emph{is not} replicated in the Jordan frame.

Finally, for natural inflation with $\zeta \in [10^{-2}, 1/6]$, violations of the first and third stability conditions at the beginning of inflation indicate a phase transition. This underscores the high sensitivity of the model's thermodynamic phase structure to the non-minimal coupling $\zeta$. For $\zeta > 1/6$, critical points appear during the inflationary period. From a cosmological perspective, large $\zeta$ values are disfavored by observations, as the spectral index $n_s$ becomes proportional to $\zeta$ in this regime. Thus, consistency with both observational data and thermodynamic stability requires $\zeta \leq 1/6$, in quite good agreement with references suggesting couplings smaller than the conformal one \cite{Bezrukov:2008ut,Barbon:2009ya}.

Summarizing, a thermodynamic analysis of inflationary models establishes $T_{RS}$ as the physical and observable temperature of the inflaton field. This quantity decreases as inflation tends to zero and consistently matches the laws of thermodynamics, exhibiting frame-independent behavior. Conversely, in our analysis, $T_{RG}$ shows a counter-intuitive warming and turns out to be nonphysical. The thermodynamic formalism reveals that the inflaton's behavior might have phase transitions and critical points, directly linking the dynamics of inflation to them.

The specific thermodynamic outcomes are highly model-dependent. For large field models, including Starobinsky, E-model, and T-model potentials, a persistent violation of a key stability condition indicates a second-order phase transition throughout the inflationary epoch, which we interpret as the mechanism that triggered its onset. The non-minimally coupled Higgs model similarly exhibits metastability and indications of a phase transition at the beginning of inflation in both Jordan and Einstein frames, reinforcing the picture of inflation as a process initiated by a thermodynamic phase transition.

Among small field models, hilltop potentials are deeply metastable at the Planck epoch and exhibit a critical point near the end of inflation, suggesting a supercritical fluid state. However, incorporating quantum corrections via a Coleman-Weinberg potential alters this picture, causing the critical point to vanish and a second order phase transition to appear. Hybrid inflation is thermodynamically stable, while natural inflation features a wide metastable region, allowing for a phase transition that could also trigger inflation.
When non-minimal coupling is introduced, the critical point vanishes for the hilltop potentials.

A similar behavior is observed for hybrid inflation along the valley, where the critical point disappears even at the critical parameter value $\mu/M_{pl} = 1/\sqrt{2}$. For these small-field models, a phase transition is expected to occur just before the beginning of inflation, $\Delta N > 60$, which we interpret as the physical mechanism that initiated the inflationary epoch.
It was also found that $\zeta$ has a maximum value for the thermodynamic description to be \emph{frame-independent}, which matches the value $1/6$ where the theory becomes conformally invariant. Finally, for natural inflation, the thermodynamic outcome depends critically on the coupling constant. For $\zeta = 10^{-4}$, no stability conditions are violated. However, for $\zeta \geq 10^{-2}$, the behavior changes and a phase transition occurs near the onset of inflation. As in the case of small-field models, this can be interpreted as the origin of inflation.

\section{Outlooks and perspectives} \label{section5}

We here presented a thermodynamic analysis for scalar field inflation investigating two distinct definitions of temperature, namely a ``real gas'' fluid-like temperature $T_{RG}$, which causes the system's thermodynamic EoS to mimic those of a real gas, and a ``relativistic species-like'' temperature $T_{RS}$, which reflects the scale factor dependence characteristic of relativistic species in the early universe.

We found that $T_{RS}$ emerges as the most physically-consistent choice, in view of its monotonically decreasing behavior throughout the inflationary epoch, matching the expected universe cooling during the accelerated expansion. Moreover, the associated fundamental equation for the entropy is well-defined and satisfies all laws of thermodynamics, reaching a minimum as the temperature approaches zero.

Conversely, we showed that the real gas temperature $T_{RG}$ increases during inflation. This, along with a possible negative heat capacity\footnote{For applications of this property in cosmology, see e.g. \cite{Lynden-Bell:1998pzh,Einarsson:2004cf,Luongo:2012dv,DAgostino:2024ymo}.}, renders it an unstable and non-viable description for the inflaton field.

A central feature of our formalism was presented analyzing the thermodynamic description in  non-minimal scenarios. We explored a possible \emph{frame dependence}, i.e., the consistency of thermodynamic behavior between the Jordan and Einstein frames, in terms of the value of the coupling constant, $\zeta$. Precisely, for couplings $\zeta \leq 1/6$, the thermodynamic inflationary behavior  does not depend on the frame, Einstein or Jordan, exhibiting a \emph{frame independence}. There, the onset of phase transitions and the (non-)existence of critical points are consistent between the frames and physically robust.

Nevertheless, this property breaks down for $\zeta > 1/6$, where the Einstein frame potential develops divergences, quite in disagreement with the Higgs inflation prescription \cite{Rubio:2018ogq}, albeit in line with other findings, see e.g. \cite{Sakai:1998rg,Gao:2025onc}.

We thus showed that this occurrence establishes a clear thermodynamic bound on the coupling constant, with $\zeta \leq 1/6$ emerging as necessary conditions for a frame-independent thermodynamic description.

On the other hand, our thermodynamic approach has been used to distinguish between inflationary models favored by current cosmological bounds, suggesting a possible \emph{alternative criterion} toward selecting the most viable inflationary model,  complementing \emph{de facto} the standard comparison based on $n_s$ and $r$, from the  Planck satellite constraints \cite{Plank2018inflation}.

In particular, we remarked that the large-field models favored by Planck data, i.e., the Starobinsky, E-model, and T-model potentials, consistently exhibit a thermodynamic signature of a second-order phase transition throughout the entire inflationary epoch, showing a persistent violation of stability conditions without critical points.

Conversely, we employed the hilltop potential, not fully-excluded by the Planck results, which demonstrates, instead, profound thermodynamic shortcomings, limiting its role in describing inflation.

More precisely, we found that  the classical hilltop model exhibits metastability deep in the Planck epoch where classical thermodynamics may not be applicable and, also, it features a problematic critical point near the end of inflation since it suggests that a \emph{super critical} persists after the end of inflation.

These findings have been framed into the thermodynamic issues associated with the hilltop model's observational problems, particularly related to its tension on the spectral index $n_s$, which typically requires additional fine-tuning or specific parameter choices to become viable.

The above supports the fact that our thermodynamic formalism is capable of identifying limitations, pointing out which extensions may be used as  plausible solutions. In the case of small field regime, in fact, we showed that adding a non-minimal coupling and/or quantum corrections can improve the paradigm itself, albeit this implies a subdominance of the hilltop potential with respect to the non-minimal coupling itself.

Analogously, we identified an intrinsic fine-tuning in hybrid inflation along the single-field valley, namely critical points and/or phase transitions, are reached only at the boundary of the slow-roll regime, i.e., at $\mu/M_{\rm pl} = 1/\sqrt{2}$, and in a manner that depends sensitively on the value of the coupling constant. This indicates that the model is thermodynamically fine-tuned, confirming observational issues related to it \cite{Dine:2011ws,Iso:2015wsf,Linde:2017pwt}.

Last but not least, the thermodynamic fate of natural inflation was shown to be highly sensitive to the non-minimal coupling constant $\zeta$, turning into a thermodynamic stability when the couplings are small, while for larger values undergoing a phase transition.

Accordingly, our procedure may show a direct connection between model parameters, dictated by the structure of the Lagrangian, with the underlying thermodynamics.

As future perspective, since our thermodynamic perspective seems to open novel avenues toward selecting the most viable inflationary potential, one could explore the same occurrences within the context of reheating, i.e., tracking thermodynamic entropy and temperature through the inflaton's decay. This approach could provide a novel description of the transition to a radiation-dominated universe and potentially identify phase transitions associated with specific preheating mechanisms. Moreover, incorporating \emph{quantum corrections} into the thermodynamic framework may represent another crucial  step, especially for models where such corrections qualitatively alter the classical picture, as seen in the hilltop potential. Finally, exploring potential signatures of these inflationary phase transitions in the primordial power spectrum, non-Gaussianities, or the production of primordial gravitational waves could forge a direct link between our thermodynamic formalism and future cosmological observations.

\section*{Acknowledgements}
JA appreciates the financial support provided by CONACYT through the ``Becas Nacionales para Estudios de Posgrado 2023'' with CVU 899135. UNAM and specifically the ``Posgrado en Ciencias Fisicas'' for providing the facilities to carry out this work. OL acknowledges support by the  Fondazione  ICSC, Spoke 3 Astrophysics and Cosmos Observations. National Recovery and Resilience Plan (Piano Nazionale di Ripresa e Resilienza, PNRR) Project ID $CN00000013$ ``Italian Research Center on  High-Performance Computing, Big Data and Quantum Computing" funded by MUR Missione 4 Componente 2 Investimento 1.4: Potenziamento strutture di ricerca e creazione di ``campioni nazionali di R\&S (M4C2-19)" - Next Generation EU (NGEU). The work of HQ was partially supported by by PAPIIT-DGAPA-UNAM, grant No. 108225, and by CONAHCYT,  grant No. CBF-2025-I-253.

\onecolumngrid

\clearpage

\appendix
\section{Plots} \label{appendix plots}
\onecolumngrid
This appendix contains supporting plots referenced throughout the main text. These figures provide additional detail or context for the results discussed in Sect \ref{section4}

\begin{figure}[!htb]
    \includegraphics[width = 0.90\linewidth]{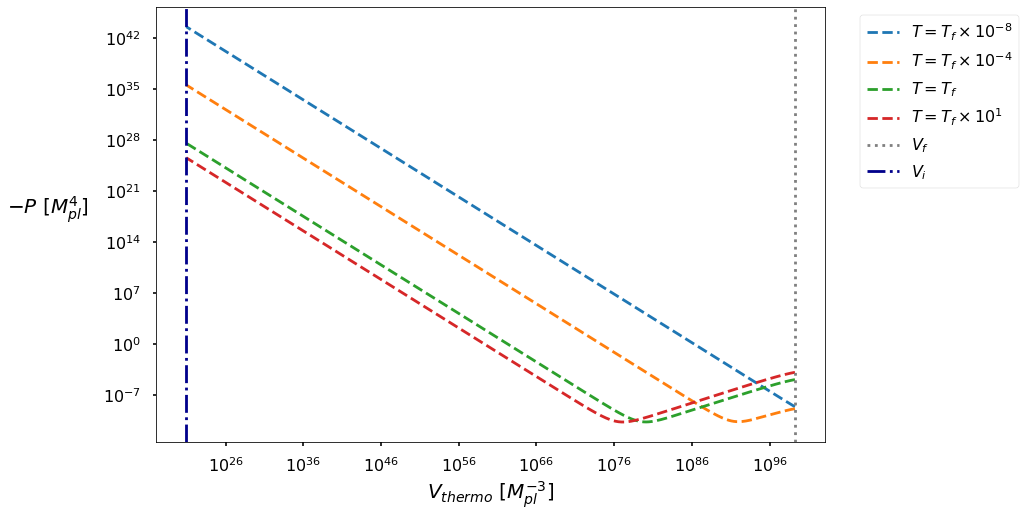}\\
    \vspace{5mm}
    \includegraphics[width = 0.90\linewidth]{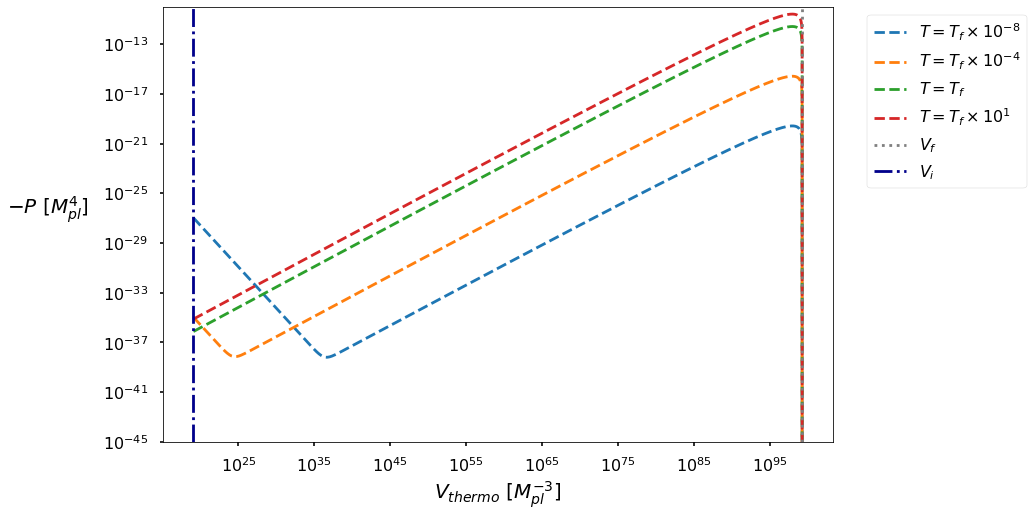}
\caption{Isotherms of the relativistic species-like temperature for different temperature values, presented on a log-log scale. The blue dashed-dotted vertical line indicates the volume $V$ at the beginning of inflation (assuming $60$ e-folds), while the black dotted vertical line indicates the volume $V$ at the end of inflation. The top panel shows the non-minimally coupled potential in the Einstein frame, and the bottom panel shows it in the Jordan frame. The isotherms in both frames exhibit a shape characteristic of Van der Waals fluids, suggesting that thermodynamic phase transitions may occur.}
\label{fig: isotherm comparison 2 models}
\end{figure}

\end{document}